\documentclass[aps,preprint,amsmath,amssymb,nofootinbib]{revtex4-1}
\usepackage{xcolor}
\usepackage{graphicx} 
\usepackage[caption=false]{subfig}

\usepackage{gensymb} 
\usepackage{amsmath} 
\usepackage{braket} 
\usepackage{booktabs} 
\usepackage{ulem}
\usepackage[colorlinks=true]{hyperref} 
\hypersetup{colorlinks,
	citecolor=blue
}
\usepackage[capitalize]{cleveref}

\newcommand{\ba}{\begin{eqnarray}}
\newcommand{\ea}{\end{eqnarray}}

\begin{document}


\title{Analysis of NOvA and MicroBooNE charged-current inclusive neutrino measurements within the SuSAv2 framework}


\author{J. Gonzalez-Rosa$^a$, G. D. Megias$^a$, J. A. Caballero$^{a,b}$, M. B. Barbaro$^{c,d}$}
\affiliation{$^a$Departamento de F\'isica At\'omica, Molecular y Nuclear, Universidad de Sevilla, 41080 Sevilla, Spain}
\affiliation{$^b$ Instituto de F\'{\i}sica Te\'orica y Computacional Carlos I, Granada 18071, Spain}

\affiliation{$^c$Dipartimento di Fisica, Universit\`a di Torino, Via P. Giuria 1, 10125 Torino, Italy}
\affiliation{$^d$INFN, Sezione di Torino, Via P. Giuria 1, 10125 Torino, Italy}



\date{\today}

\begin{abstract}
In this work we compare the SuSAv2 model, based on the superscaling phenomenon and the relativistic mean field theory, with charged-current inclusive neutrino cross sections from the NOvA and MicroBooNE experiments, whose targets are composed primarily by $^{12}$C 
and $^{40}$Ar, respectively. The neutrino energy in these experiments covers a kinematic range from tens of MeV to roughly 20 GeV. Thus, we consider the different reaction mechanisms that contribute significantly to these kinematics, namely quasielastic, two-particle two-hole meson exchange currents, resonances and deep inelastic scattering contributions. 
\end{abstract}


\maketitle
\section{Introduction}\label{Introduction}



Neutrino scattering processes and their detection through nuclear interactions play a crucial role nowadays in uncovering key aspects of physics, such as charge-parity (CP) violation and the associated matter-antimatter asymmetry, the dynamics of supernovae, and the determination of the neutrino mass hierarchy \cite{alvarez-ruso_nustec_2018}. The uncertainties in these nuclear interactions are of paramount relevance in the study of these topics. Therefore, the development of neutrino interaction models like the one presented in this work and its subsequent implementation in experimental event generators are essential for the success of neutrino experiments. 

In the context of neutrino oscillation analyses, various long-baseline neutrino facilities are presently running, or have been completed in past years. They operate at different neutrino energies, ranging from tens of MeV to several GeV, peaking their fluxes at around 0.6-0.8 GeV in the case of MiniBooNE~\cite{miniboone_collaboration_first_2010}, MicroBooNE~\cite{abratenko_first_2019-1}, or T2K~\cite{t2k_collaboration_measurement_2018}, but also at higher energies - roughly from 2 to 6 GeV - in MINERvA~\cite{minera_collaboration_double-differential_2020} or NOvA~\cite{acero_measurement_2023}. Future experiments, such as DUNE~\cite{dune_collaboration_long-baseline_2016} and Hyper-Kamiokande~\cite{proto-collaboration_hyper-kamiokande_2018}, will also be able to explore larger neutrino energies. The characterization of neutrino oscillation properties in these experiments depends on reconstruction methods based on the final-state particles detected after the reactions of neutrinos with the targets in the detectors. This process relies on Monte Carlo event generators~\cite{GENIE:2021npt,JUSZCZAK2006211,Hayato:2009zz,PhysRevD.107.033007,GARDINER2021108123,soplin2024montecarlosimulationdevelopment} that simulate the experimental conditions and the nuclear models implemented in them. The role of the different nuclear reaction mechanisms in these neutrino interactions is largely dependent on the neutrino energy range. In the domain from a few MeV to a few GeV, the quasielastic (QE) regime is very prominent. This regime is very important in the MicroBooNE or T2K experiments and is characterized by processes having one nucleon knocked out in the final state. Another relevant contribution in this domain is the emission of two nucleons, corresponding to the excitation of two-particle two-hole (2p2h) states induced by meson-exchange currents (MEC). This process is sometimes also indicated by MEC or 2p2h-MEC. As the energy transferred from the neutrino to the target increases, the nucleons can be excited forming resonances which rapidly decay, emitting pions or other mesons. This regime is called the resonance region (RES) where at similar kinematics there are also other inelasticities linked to non-resonant meson production. At larger energies, the probe starts to interact with the quarks inside the nucleons, opening the deep-inelastic scattering (DIS) channel. These high-energy contributions are of particular relevance for MINERvA or NOvA and also for future experiments like DUNE \cite{dune_collaboration_long-baseline_2016} and Hyper-Kamiokande \cite{proto-collaboration_hyper-kamiokande_2018}. 
 
Measurements in neutrino experiments are diverse, focusing on various channels classified on the basis of the observed final states.  ``CC-inclusive" measurements involve detecting only the final lepton in charged-current (CC) reactions, with all of the aforementioned channels contributing to the cross section. The ``CC1pi" channel refers to events in which a single pion is observed in the final state.  Other common measurements include ``CC0$\pi$" and ``semi-inclusive" channels. The CC0$\pi$ channel involves CC events where no pions are detected in the final state. This cross section is expected to be dominated by QE scattering and 2p2h contributions. However, if a resonance is produced and the pion from its decay is subsequently reabsorbed by the nucleus, an event with the same topology will still be detected. Therefore, pion absorption effects must be accounted for in CC0$\pi$ measurements. In semi-inclusive measurements, one or more protons and/or other hadrons are detected in coincidence with the lepton, providing access to the hadrons' kinematics. These measurements are particularly sensitive to the nuclear modeling incorporated into theoretical calculations. 

In this work, we focus on CC-inclusive measurements from the NOvA and MicroBooNE experiments, having analyzed CC0$\pi$ and semi-inclusive processes in previous works~\cite{PhysRevC.102.064626,PhysRevD.109.013004,PhysRevD.106.113005,megias_charged-current_2016}, and we compare these data with predictions for the QE and inelastic channels from the SuSAv2 model~\cite{gonzalez-jimenez_extensions_2014,megias_inclusive_2016,gonzalez-rosa_susav2_2022} , which is based on the Relativistic Mean Field (RMF) theory and the superscaling phenomenon, and that has been partially implemented in the event generator GENIE~\cite{PhysRevD.103.113003,PhysRevD.101.033003}. The 2p2h channel is obtained following the RFG-based calculations of~\cite{Simo_2017,AMARO2011151}. More details about the theoretical description of the different reaction mechanisms and their main features are given in the next section. Note also that the nuclear targets analyzed in this work differ from the carbon-based ones from other experiments such as T2K or MINERvA. The NOvA near detector, located at Fermilab, is aligned with the NuMI neutrino beam. This beam interacts with a target composed of 67\% carbon, 16\% chlorine, 11\% hydrogen, 3\% titanium, and 3\% oxygen, along with trace amounts of other elements \cite{PhysRevLett.130.051802,NOvAonline}. The NOvA experiment also employs a far detector but it will not be the object of this study. Moreover, the MicroBooNE experiment \cite{MicroBooNEonline}, also located at Fermilab, operates on the Booster Neutrino Beamline and uses a liquid argon time projection chamber (LArTPC) as its detector, with a target primarily composed of argon. The MicroBooNE's argon-based detector provides detailed insights into the interaction properties of the neutrinos, while the NOvA detectors' carbon-rich targets are particularly relevant for oscillation and CP violation studies.

In what follows, we present the theoretical formalism in Section~\ref{Theoretical}, describing the features of the different reaction mechanisms and detailing the concept of superscaling within the SuSAv2 framework. In Section~\ref{Results}, we show and discuss our theoretical predictions in comparison with CC-inclusive neutrino cross section data, firstly for NOvA (Sect.~\ref{NOvA}) 
and later for MicroBooNE (Sect.~\ref{MicroBooNE}). Finally, we draw our conclusions in 
Section~\ref{Conclusion}.

\section{Theoretical Background}\label{Theoretical}


In this work, the comparison with measurements from the experiments mentioned above will be carried out using the SuSAv2 model for the QE and inelastic regimes together with RFG-based 2p2h calculations from~\cite{Simo_2017,AMARO2011151}. The SuSAv2 model~\cite{gonzalez-jimenez_extensions_2014,megias_charged-current_2017,megias_inclusive_2016,megias_charged-current_2016} is based on the Relativistic Mean Field 
theory~\cite{RING1996193,gonzalez-jimenez_nuclear_2019} and the superscaling phenomenon exhibited by the large amount of inclusive lepton-nucleus scattering data~\cite{donnelly_superscaling_1999-2}. In the RMF theory, the bound and scattered nucleon wave functions are solutions of the Dirac-Hartree equation in the presence of energy-independent real scalar and vector potentials, with parameters fitted to the saturation properties of nuclear matter. On the other hand, superscaling is related to the general behavior observed in the nuclear response of different nuclear targets to a leptonic probe 
for different values of the momentum transferred to the nucleus.
Specifically, if the double-differential inclusive lepton-nucleus cross section, with respect to the energy transfer $\omega$ and the solid scattering angle $\Omega$, is divided by an appropriate single-nucleon cross section, it becomes independent on both the transferred momentum $q$ and the nuclear species, the latter being characterized by the Fermi momentum $k_F$. This yields a dependence on a single variable, $\psi$, known as the scaling variable. This property is satisfied when $q$ exceeds approximately 400 MeV, namely in a region where low-energy nuclear effects are not prominent. 

In the SuperScaling Approach (SuSA), the cross section can thus be expressed as the product of a single-nucleon cross section and a scaling function $f(\psi)$, which encapsulates information about the nuclear structure and dynamics. This factorization is assumed to hold across the entire energy spectrum, covering various nuclear processes, from quasielastic to deep inelastic scattering, each process being associated with a different single-nucleon function. In the first SuSA approach~\cite{amaro_using_2005}, the scaling function was extracted from quasielastic electron scattering ($e,e^{\prime}$) data as 

\begin{equation}
f(\psi)=\frac{\frac{d^{2}\sigma}{d \Omega_{e} d \omega}}{\sigma_{Mott}(V_{L}G^{ee^{\prime}}_{L} + V_{T}G^{ee^{\prime}}_{T} ) } ,
\end{equation}
where $\sigma_{Mott}$ is the Mott cross section, $V_{L,T}$ are the leptonic kinematic factors and $G^{ee^{\prime}}_{L,T}$ are the elastic single-nucleon responses dependent on the electric and magnetic form factors of the nucleon.
%
%
A more detailed description of the superscaling approach can be found in \cite{amaro_using_2005,amaro_electron-_2020-1, amaro_neutrino-nucleus_2021-1, megias_inclusive_2016, megias_charged-current_2017, megias_neutrinooxygen_2018-1, megias_analysis_2019-2}. 
This general feature enables to extract experimental information about the nuclear dynamics in these processes and it should also be reproduced by all theoretical nuclear models. In the case of the RMF model, the superscaling behavior is well fulfilled, reproducing the experimental scaling data from electron reactions as well as electron scattering data in general. Thus, an improved version of the SuSA model, called SuSAv2, was developed for the quasielastic region, using the scaling functions obtained from RMF and relativistic plane wave impulse approximation (RPWIA) calculations~\cite{gonzalez-jimenez_extensions_2014,megias_inclusive_2016} for electron and neutrino processes. Within the SuSAv2 framework, the information from RMF is used to obtain a complete set of scaling functions that permit to reproduce the complex RMF microscopic calculations in a straightforward formalism and that embodies the nuclear dependence of lepton-nucleus interactions.
The SuSAv2 model, originally developed for the quasielastic (QE) regime, was subsequently developed for the inelastic regime for electrons~\cite{megias_inclusive_2016} and later for neutrinos~\cite{gonzalez-rosa_susav2_2022,PhysRevD.108.113008}. In these works, this approach has yielded an overall agreement with ($e,e'$) data and with MicroBooNE, T2K, ArgoNEUT and MINERvA measurements.

Within this framework, and for charged-current (CC) quasielastic neutrino-nucleus scattering, the differential cross section can be written in the general form:
\begin{widetext}
\begin{equation}
    \frac{d^2\sigma}{d\Omega d p_{l}}=\sigma_{0}(V_{CC}R_{CC} +2V_{CL}R_{CL} + V_{LL}R_{LL} + V_{T}R_{T} + 2V_{T^{\prime}}R_{T^{\prime}})
\end{equation}
\end{widetext}
in terms of the $\sigma_0$ factor, the leptonic kinematic factors ($V_{K}$), and the nuclear response functions ($R_{K}$) that depend on the elastic single-nucleon response functions and the scaling functions as detailed in~\cite{PhysRevD.108.113008}, and where $p_l$ is the three-momentum of the outgoing lepton.



Moreover, in the inelastic regime, the nuclear responses are written in terms of the single-nucleon inelastic structure functions, $G^{inel}_{K}$, and of the scaling function, $f(\psi)$, as follows: 

\begin{widetext}
\begin{equation}\label{RKinel}
R^{inel}_{K}(q,\omega)=N\,\frac{2T_Fm_{N}^{3}}{k^{3}_{F}q}\,\int_{\mu_{X}^{min}}^{\mu_{X}^{max}}d\mu_{X}\mu_{X}f^{\rm SuSAv2}(\psi_{X})\, G^{inel}_{K},
\end{equation}
\end{widetext}
being $N$ the number of nucleons participating in the reaction, $k_{F}$ the Fermi momentum, and $T_{F}\equiv \sqrt{m_N^2 + k_{F}^{2}} - m_N$ the Fermi kinetic energy. $f^{\rm SuSAv2}$ is the SuSAv2 scaling function, 
 $\mu_{X}$ is the reduced invariant mass ($\mu_{X}=W_{X}/m_{N}$) and $\psi_X$ the generalization of the scaling variable $\psi$ from the QE to the inelastic regime as shown in~\cite{gonzalez-rosa_susav2_2022}. 
Depending on the limits of the integral \eqref{RKinel} and on the inelastic structure functions $G^{inel}_{K}$ employed, we can focus on the different channels that contribute to the full inelastic regime, such as resonance production or DIS. In the case of the resonance regime, we have recently incorporated the Dynamical Coupled-Channels (DCC) model from the Osaka group~\cite{nakamura_dynamical_2015-3,Nakamura:2018ntd, DCConline} into our framework in the so-called SuSAv2-DCC approach~\cite{PhysRevD.108.113008}, which is valid in the region  $W_{X}\le 2.1$ GeV  and $Q^{2} \le 3$ GeV~\cite{nakamura_dynamical_2015-3}. The DCC model provides a very accurate description of the nucleon resonances based on extensive analyses from ANL data. 
 It is also worth mentioning that the description of single-nucleon inelastic structure functions in most resonance models is based on either phenomenological fits of experimental data that account for the different nucleon resonances and other effects or extrapolations of QCD results to low-mid kinematics. The most advanced approaches in this domain are the above-mentioned DCC model \cite{nakamura_dynamical_2015-3,Nakamura:2018ntd, DCConline} and the MK  \cite{Kabirnezhad_2018,Kabirnezhad:2022znc} approach. 
 Moreover, the description of nuclear dynamics in these resonance models varies and goes from simple RFG-based approaches to more microscopic descriptions such as pion-production RMF models~\cite{gonzalez-jimenez_pion_2018-1,gonzalez-jimenez_nuclear_2019}. Recently, there have also been efforts from theoretical groups to improve descriptions of nucleon distortions in lepton-induced single-pion production via microscopic calculations~\cite{PhysRevC.109.024608}.

In the SuSAv2-inelastic model, deep inelastic scattering contributions are considered as processes not taken into account by the DCC approach below or above $W_{X}=2.1$ GeV. In this case, we define the DIS contributions above the resonance region described by DCC as ``TrueDIS" and within this resonance region as ``SoftDIS". For TrueDIS, the limits are $W_{X}^{min}=2.1$ GeV and $W_{X}^{max}=m_N + \omega - E_{s}$, with $E_{s}$ being the separation energy. The SoftDIS contribution shares the kinematical limits of the DCC model and is obtained as the SuSAv2-inelastic result minus the SuSAv2-DCC one. More details can be found in \cite{PhysRevD.108.113008}. 

For the DIS regime, together with the SuSAv2 scaling functions, we can employ parton distribution functions (PDFs)~\cite{callan_high-energy_1969-1,tooran_qcd_2019-2, stein_electron_1975-2, gluck_dynamical_1998-1} or phenomenological single-nucleon structure functions derived from fits to electron scattering data, such as those from the Bosted-Christy or Bodek-Ritchie parameterizations~\cite{bodek_axial_2013-1, bodek_fermi-motion_1981-1, bodek_further_1981-1, bodek_experimental_1979-2, liang_notitle_2004, bosted_empirical_2008-1}. 
In this work, the single-nucleon inelastic structure functions for the DIS contributions are based on the Bodek-Ritchie (BR) parameterization for consistency with our previous article \cite{PhysRevD.108.113008}, because PDFs do not work well at $Q^{2}$ below 0.8 GeV$^{2}$ and Bosted-Christy (BC) is not suitable for high kinematics ($\omega \gtrsim 10$ GeV). Nevertheless, at MicroBooNE kinematics, BC results do not show noticeable differences with BR ones. For the $\nu_\mu$ NOvA case, where the flux goes up to 20 GeV, differences up to 20\% at very forward angles can be observed for the SoftDIS channel, being BC larger. However, for the TrueDIS channel, which explores higher kinematics, the limitations in the BC approach reduce its contribution, compensating the increase in the SoftDIS channel, which eventually leads to similar results with regard to BR. Note also that unlike other channels, deep-inelastic scattering contributions are very dependent on the flux high-energy tail.
For the resonance regime, we employ the DCC functions in combination with the SuSAv2 scaling functions, which, together with the DIS contributions, were successfully applied for the analysis of electron scattering data as well as for CC-inclusive T2K and MINERvA measurements on carbon targets~\cite{PhysRevD.108.113008}. 

In this work, we will continue these previous analyses by testing the full SuSAv2 model with CC-inclusive MicroBooNE~\cite{PhysRevLett.128.151801, microboonecollaboration2024measurementthreedimensionalinclusivemuonneutrino} and NOvA~\cite{PhysRevD.107.052011, PhysRevLett.130.051802} data on different targets. 
The contributions considered for the different nuclear reaction channels in the Results section, with their corresponding acronyms, are summarized in Table~\ref{Table} as well as the model used to describe them.


\begin{table*}
\begin{tabular}{|c | c | c|}
\hline
\hline
Acronym & Definition of the reaction channel & Model \\
\hline
 QE    & Quasielastic &  SuSAv2 {QE} \\
  MEC   & 2p2h MEC excitations   & RFG-MEC  \\
  RES  &   Resonant  &   SuSAv2-DCC \\
  SoftDIS &  Deep inelastic scattering  &  SuSAv2 inelastic $-$  SuSAv2-DCC  \\
   & and non-resonant ($W_{X}<2.1$ GeV) & \\
  TrueDIS &  Deep inelastic scattering ($W_{X}>2.1$ GeV)  & SuSAv2 inelastic   \\
  \hline
  \hline
\end{tabular}
 \caption{List of acronyms used in the manuscript for the different nuclear reaction channels and the model used for each one.\label{Table}}
 \end{table*}

\section{Results}\label{Results}

In this section, we compare the NOvA and MicroBooNE CC-inclusive measurements with our predictions using the SuSAv2 model for the QE and inelastic regimes together with the RFG-MEC model as described in Table~\ref{Table}. In the following subsections, a $\chi^2$-based analysis is also presented when analyzing the results obtained.
 \subsection{NOvA} \label{NOvA}

\begin{figure*}[!htbp]
  \includegraphics[width=\textwidth]{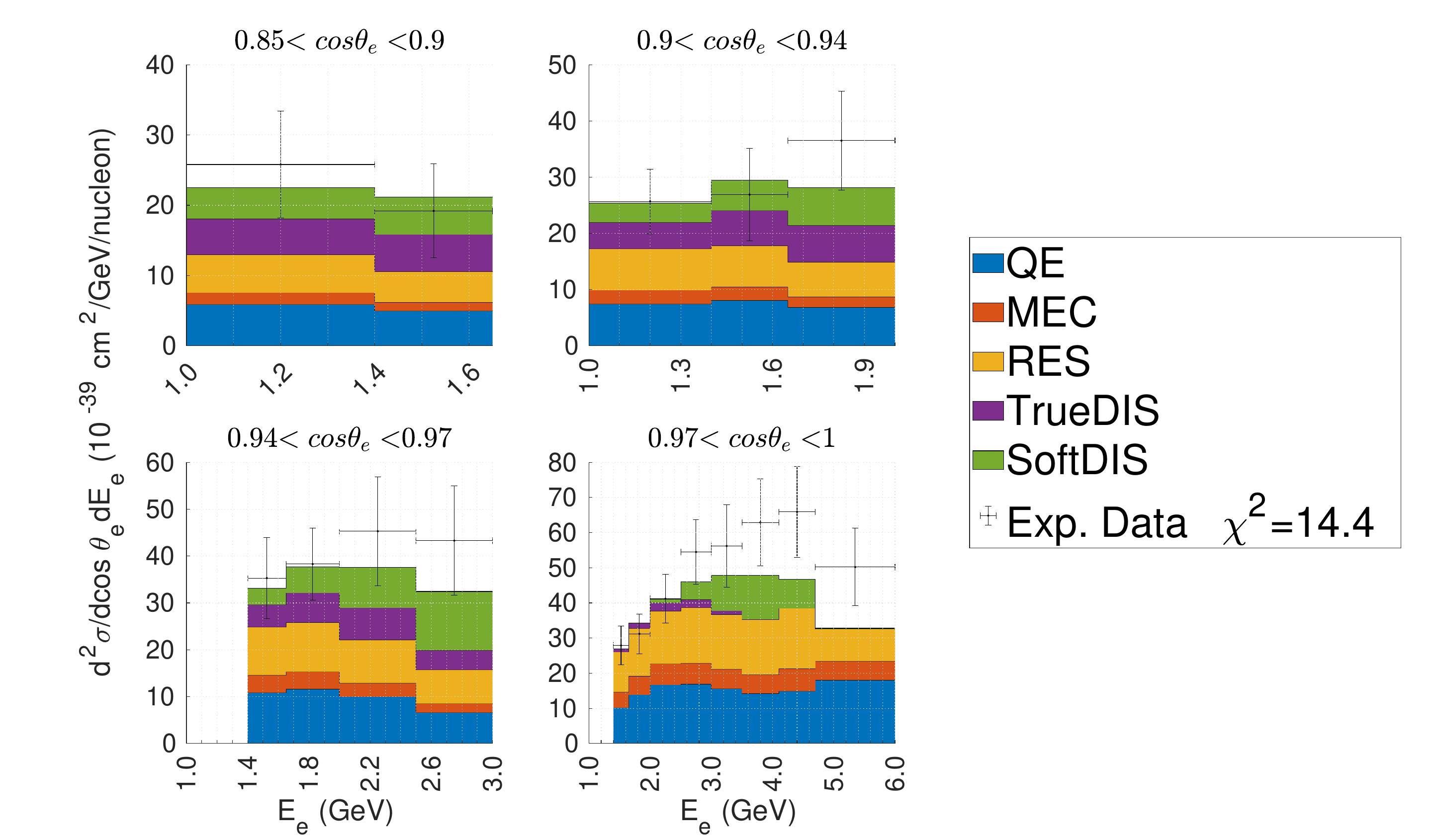}
    \caption{NOvA CC  inclusive flux-averaged double-differential cross section per target nucleon  in bins of the electron scattering angle (labeled in the panels) as a function of the electron energy. The different theoretical calculations are shown individually. Data from \cite{PhysRevLett.130.051802}. 
    \label{NOvA_Electron} }
\end{figure*}

\begin{figure*}[!htbp]
  \includegraphics[width=\textwidth]{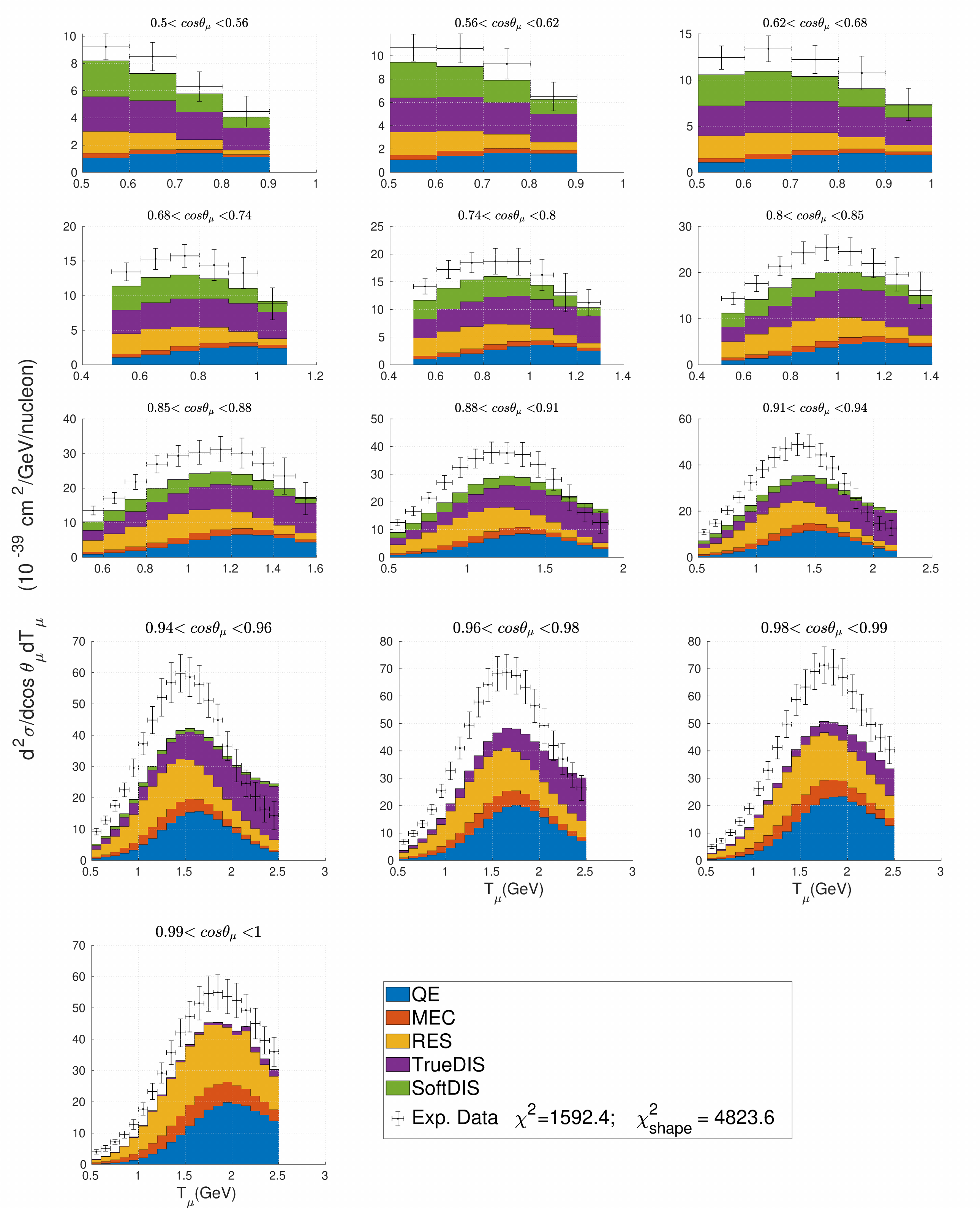}
    \caption{NOvA CC  inclusive flux-averaged double-differential cross section 
    per target nucleon  in bins of the muon scattering angle (labeled in the panels) as a function of the muon kinetic energy. The different theoretical calculations are shown individually. Data from \cite{PhysRevD.107.052011}. In the legend we show the values of $\chi^{2}$ and $\chi^{2}_{shape}=\chi^{2}F_{shape}$, where $F_{shape}$ is the only-shape factor, whose value for our predictions is 3.0. \label{NOvA_Muon}  }
\end{figure*}

In Fig.~\ref{NOvA_Electron}, we compare our models with the NOvA double-differential electron neutrino cross section for a target composed of carbon, hydrogen, chlorine, titanium, and oxygen where NOvA $\nu_e$ flux peaks around 2.4 GeV \cite{NOvAonline,PhysRevLett.130.051802}. 



In general, in the region $cos\theta_{e}<0.97$, the quasielastic regime represents roughly 25\% of the total cross section, followed by the resonance contribution with 20-25\%. The combination of the True DIS and Soft DIS channels is around 40\% of the total. In contrast, in the last plot at very forward angles, the DIS contributions fall below 15\%. Both the resonance and quasielastic channels are individually 35\% of the strength of the total result. The decrease in DIS channels at very forward angles is mainly due to the limits of the neutrino energy ($1 \leq E_{\nu} \leq 6$ GeV) which also limits the value of the energy and momentum transferred to the nucleus. 
We observe that high-energy resonances and inelasticities are produced at large transferred momentum; thus, the constraint in neutrino kinematics reduces these contributions as we explore more forward angles, i.e. when 
the available transferred energy decreases.

Our predictions tend to reproduce well the shape and the value of the experimental results apart from some underestimation at very forward angles and high electron energies which can be ascribed to some missing strength in the inelastic channels. Note also that the threshold of 6 GeV in the neutrino energy limits the inelastic contributions at high electron kinematics, as the energy transfer will not be, in general, high enough to produce resonance and other inelasticities in a significant way. Nevertheless, our $\chi^{2}$-value is a bit smaller 
than the one from other models used in generators such as NuWro, GiBUU or GENIE~\cite{PhysRevLett.130.051802}.




In Fig.~\ref{NOvA_Muon}, the NOvA double-differential muon neutrino cross section is represented for the same target as in the $\nu_e$ case. The average $\nu_\mu$ energy is around 4 GeV, peaking at 2 GeV \cite{NOvAonline,PhysRevD.107.052011}. Both the SuSAv2 RES and QE contributions are very similar and around 20 \% of the total cross section, being around 35 \% at very forward angles. Approximately half of the contribution comes from the DIS channels, which decrease to 20 \% at very forward angles. These large contributions, not present in the $\nu_e$ case, come from the extensive tail of the muon neutrino flux to very high energies larger than 6 GeV.  In general, our results tend to underestimate the cross section, especially in the region where the cross section peaks. In contrast, at high muon kinetic energies, we overpredict the experimental data in some cases, mostly due to the deep inelastic scattering contribution. The difference in the relevance of the different contributions in comparison with the NOvA $\nu_e$ case comes from the constraint of the tail in the neutrino flux not present in the muon data, which as commented previously have an important effect in the inelastic channels. Our $\chi^{2}$-value is comparable with the other models. Nevertheless, the shape-only factor for $\chi^{2}$ is 3, which is larger than the other Monte Carlo models. This increment is expected by observing the behavior of the deep inelastic scattering contribution. 

 \subsection{MicroBooNE}\label{MicroBooNE}

 \begin{figure*}[!htbp]
  \includegraphics[width=\textwidth]{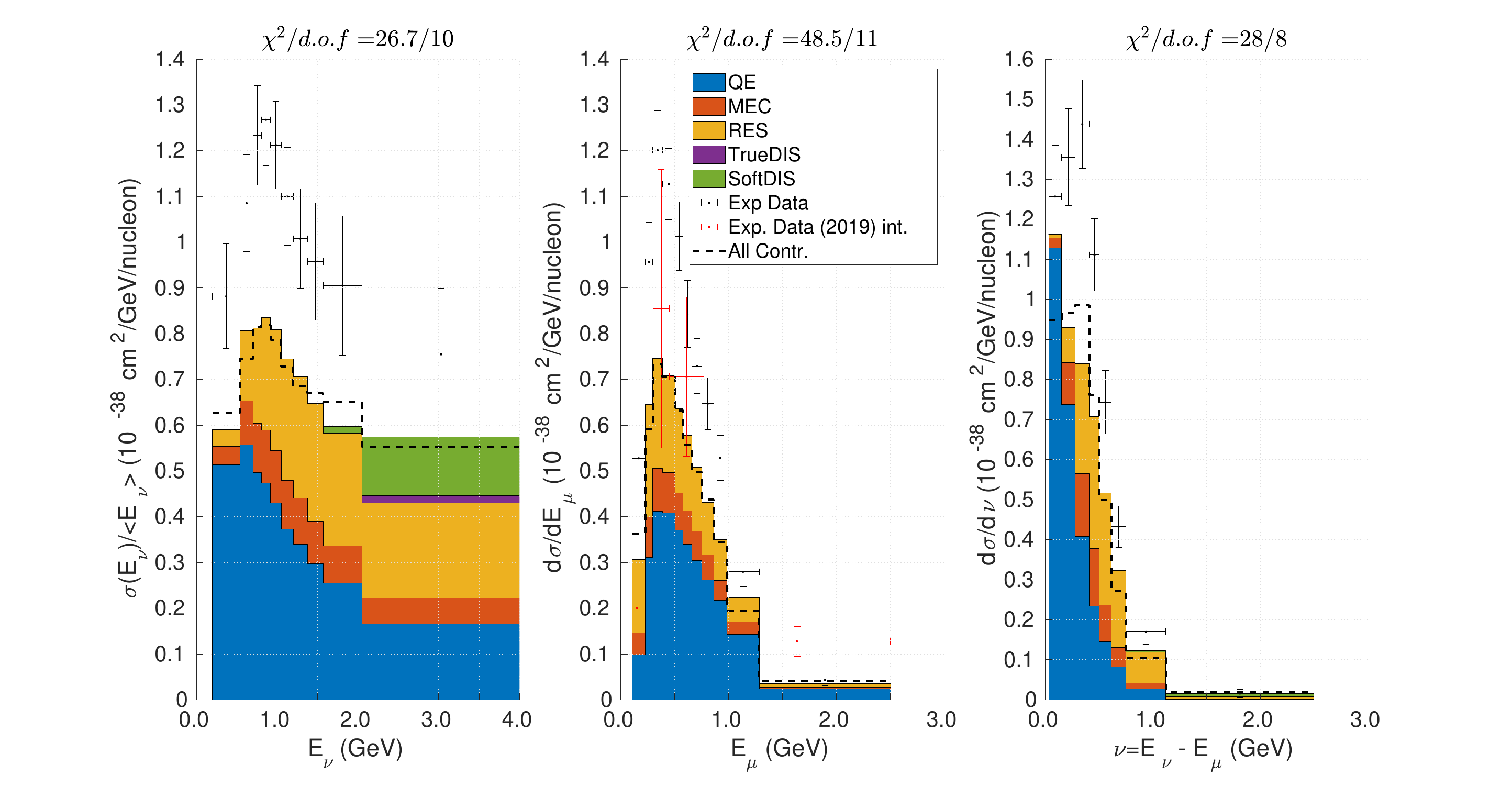}
    \caption{ (Left) MicroBooNE CC inclusive total cross section on $^{40}$Ar per target nucleon in terms of the neutrino energy. (Center) MicroBooNE CC inclusive flux-averaged single-differential cross section in terms of the muon energy. (Right) MicroBooNE CC inclusive flux-averaged single-differential cross section versus the transferred energy. Data from \cite{PhysRevLett.128.151801}.\label{MicroBooNE_Single} }
\end{figure*}




In Fig.~\ref{MicroBooNE_Single}, we show the MicroBooNE $\mu_\nu$ CC inclusive total and differential cross sections using argon as a target. The flux peaks at 0.8 GeV \cite{MicroBooNEonline,abratenko_first_2019-1}, considerably lower than NOvA.

Unlike NOvA, the comparison with these MicroBooNE data requires the application of an additional smearing matrix that transforms the theoretical results with respect to the true physics quantities and that accounts for the regularization and bias of the measurement as described in~\cite{PhysRevLett.128.151801}. The application of this smearing matrix must be done to the final result. Thus, in the following plots, we show the total contribution smeared (dashed lines) together with the individual contributions for each channel, namely QE, MEC, RES, SoftDIS and TrueDIS before the application of this smearing matrix so that the net effect of this transformation can be observed.
In the left panel of Fig.~\ref{MicroBooNE_Single}, we show the total cross section weighted by the flux, in which the value of a bin that goes from $ E_{\nu}^{ini.}$ to $E_{\nu}^{end}$  is calculated using the following expression:

\begin{equation}
    <\sigma>=\frac{1}{\int^{E_{\nu}^{end}}_{E_{\nu}^{ini.}}dE_{\nu}\phi(E_{\nu})}\int_{E_{\nu}^{ini.}}^{E_{\nu}^{end}}dE_{\nu}\int_{-1}^{+1}{dcos \theta}\int{dE_{\mu}\frac{d^{2}\sigma}{dcos\theta dE_{\mu}}\phi(E_{\nu})},
\end{equation}
being $\phi(E_{\nu})$ the neutrino flux. We have observed that this result is very similar to the one obtained for the total $\nu_\mu$-Ar cross section without considering any flux. Only small differences appear for the most extreme $E_\nu$ bins which are due to the rapid increase or decrease of the flux for that particular kinematics.

In the middle and right panels of Fig.~\ref{MicroBooNE_Single} we show the MicroBooNE flux-averaged single differential cross sections as a function of the muon and transferred energies, respectively. In all panels, a similar underestimation of the data is noticed that can be ascribed to some missing strength in the inelastic channels, although this effect is not observed in a previous work~\cite{PhysRevD.108.113008} compared to other experiments at similar kinematics.

In general, non-inelastic contributions are around 40 \% for the total cross section, around 65\% for the single-differential ones. On the other hand, the resonance channel gives 36 \% of the strength of the total cross section and 30\% for the single-differential cross section. The underestimation is also exhibited by GENIE, NEUT and other Monte Carlo simulations~\cite{PhysRevLett.128.151801}. Note also that the MicroBooNE data~\cite{PhysRevLett.128.151801} used in these plots exhibit an increase with regard to previous MicroBooNE measurements\cite{abratenko_first_2019-1} where our models produced a good comparison with the data. As we can observe, the total cross section measured by MicroBooNE is larger than our prediction.
In contrast with these results, our previous analysis~\cite{PhysRevD.108.113008} of T2K data on a hydrocarbon target at similar kinematics showed a good description of data in analogy to Monte Carlo simulations.

 \begin{figure*}[!htbp]
  \includegraphics[width=\textwidth]{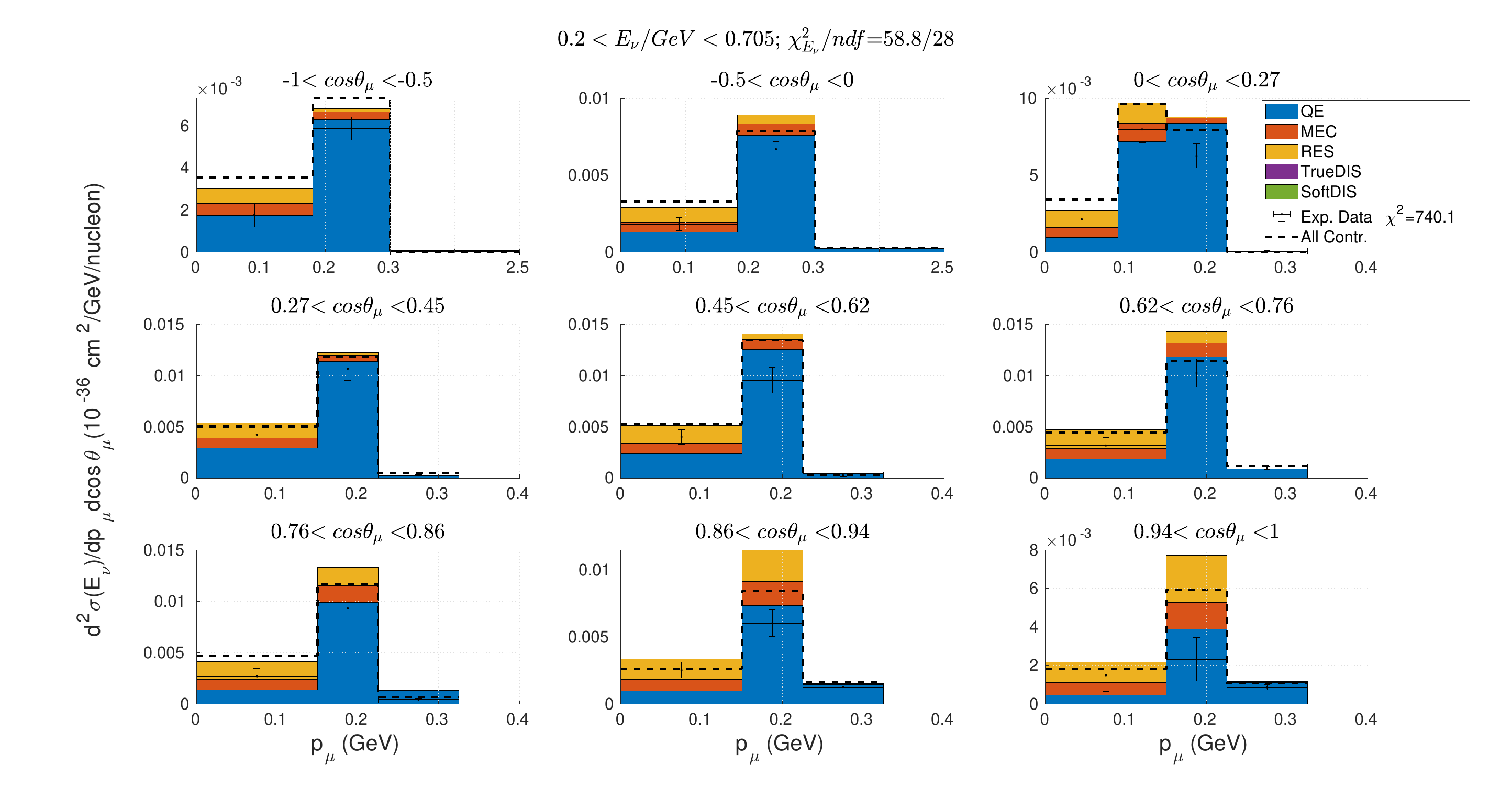}
    \caption{MicroBooNE CC inclusive flux-averaged differential 
    cross section on $^{40}$Ar per target nucleon  in bins of the muon scattering angle (labeled in the panels) as a function of the muon momentum for the neutrino energy bin of 0.2-0.705 GeV. The different contributions are shown individually before the application of the smearing matrix. Data from \cite{microboonecollaboration2024measurementthreedimensionalinclusivemuonneutrino}.
    \label{MicroBooNE_Triple_1} }
\end{figure*}

 \begin{figure*}[!htbp]
  \includegraphics[width=\textwidth]{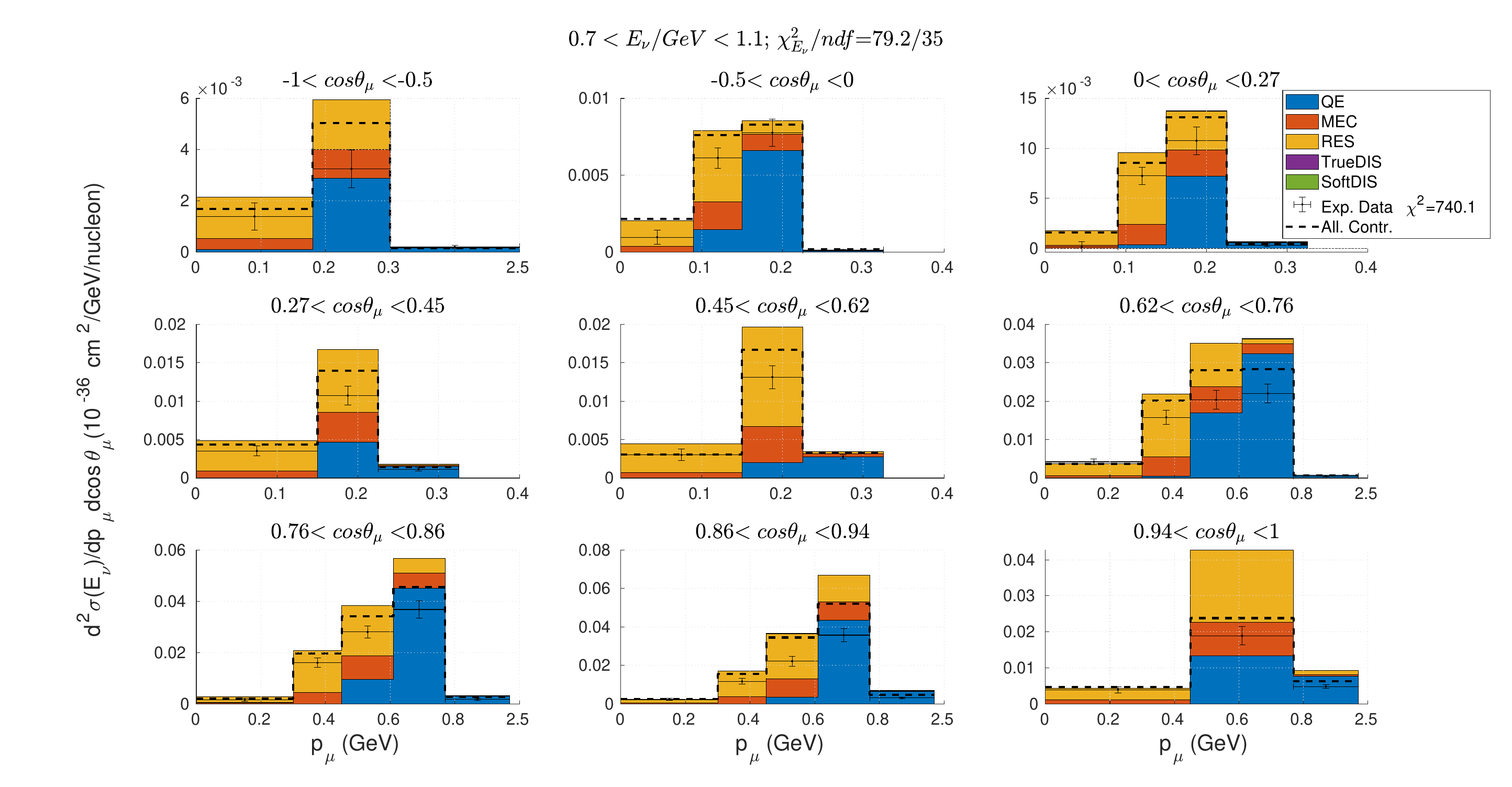}
    \caption{MicroBooNE CC  inclusive flux-averaged differential cross section on $^{40}$Ar per target nucleon  in bins of the muon scattering angle (labeled in the panels) as a function of the muon momentum for the neutrino energy bin of 0.7-1.1 GeV.  The different contributions are shown individually before the application of the smearing matrix. Data from \cite{microboonecollaboration2024measurementthreedimensionalinclusivemuonneutrino}.
    \label{MicroBooNE_Triple_2} }
\end{figure*}

 \begin{figure*}[!htbp]
  \includegraphics[width=\textwidth]{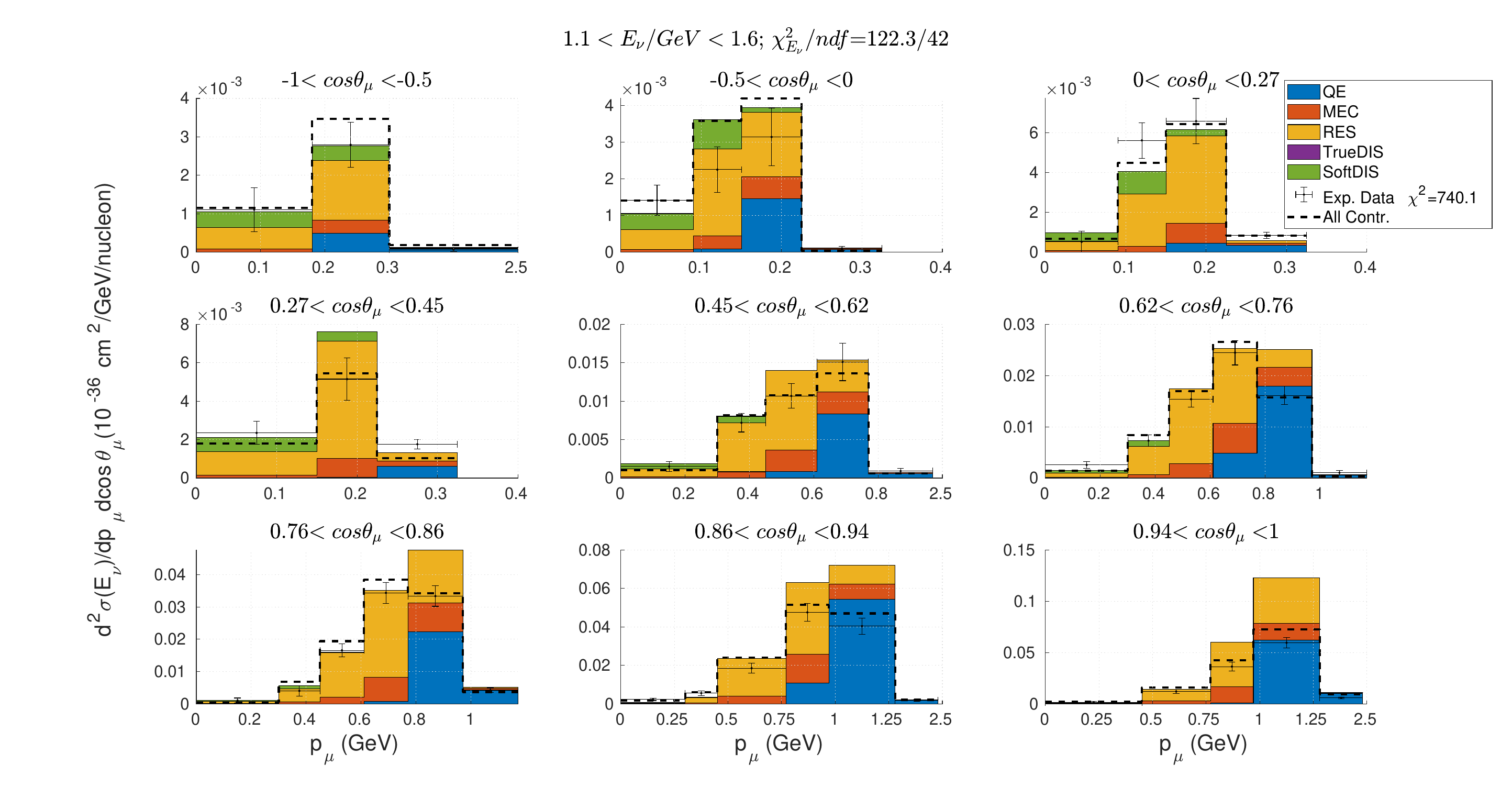}
    \caption{MicroBooNE CC  inclusive flux-averaged differential cross section on $^{40}$Ar per target nucleon  in bins of the muon scattering angle (labeled in the panels) as a function of the muon momentum showing the neutrino energy bin of 1.1-1.6 GeV.  The different contributions are shown individually before the application of the smearing matrix. Data from \cite{microboonecollaboration2024measurementthreedimensionalinclusivemuonneutrino}.
    \label{MicroBooNE_Triple_3} }
\end{figure*}

 \begin{figure*}[!htbp]
  \includegraphics[width=\textwidth]{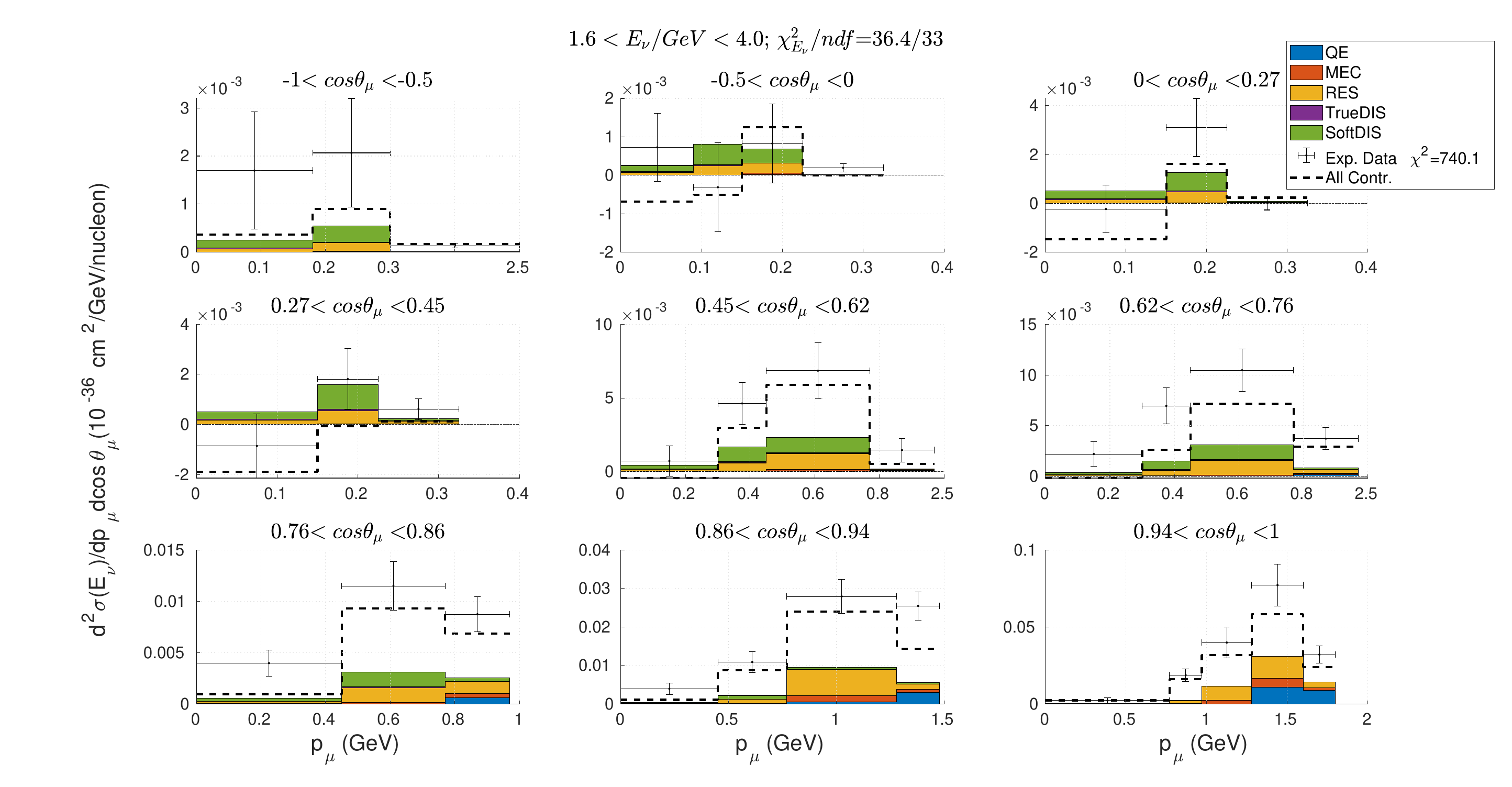}
    \caption{MicroBooNE CC  inclusive flux-averaged differential cross section on $^{40}$Ar per target nucleon  in bins of the muon scattering angle (labeled in the panels) as a function of the muon momentum showing the neutrino energy bin of 1.6-4 GeV.  The different contributions are shown individually before the application of the smearing matrix. Data from \cite{microboonecollaboration2024measurementthreedimensionalinclusivemuonneutrino}. 
    \label{MicroBooNE_Triple_4} }
\end{figure*}

 \begin{figure*}[!htbp]
  \includegraphics[width=\textwidth]{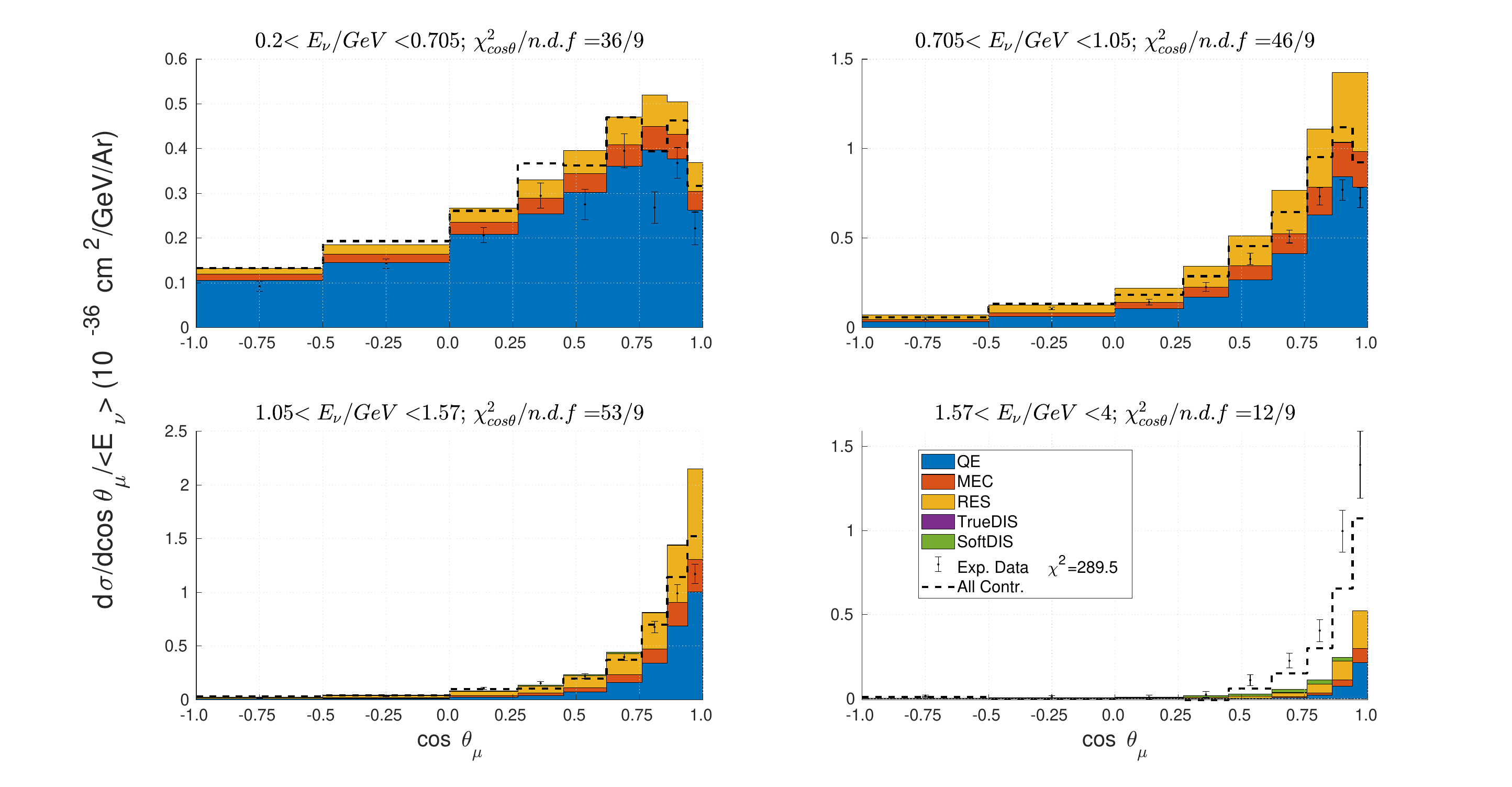}
    \caption{MicroBooNE CC  inclusive flux-averaged  differential cross section 
    on $^{40}$Ar per target nucleon  in bins of the neutrino energy (labeled in the panels) as a function of the muon scattering angle. The different contributions are shown individually before applying the smearing matrix. Data from~\cite{microboonecollaboration2024measurementthreedimensionalinclusivemuonneutrino}.
    \label{MicroBooNE_Triple_costh} }
\end{figure*}

In Figs.~\ref{MicroBooNE_Triple_1} to \ref{MicroBooNE_Triple_4}, we show the MicroBooNE flux-averaged differential cross sections with respect to the muon momentum and scattering angle, in different bins of the neutrino energy, respectively, and using the same flux as in Fig.~\ref{MicroBooNE_Single} \cite{microboonecollaboration2024measurementthreedimensionalinclusivemuonneutrino}. In Fig.~\ref{MicroBooNE_Triple_1}, that corresponds to the lower $E_\nu$ bin, the cross section is dominated by the quasielastic contribution which is still very important in Fig.~\ref{MicroBooNE_Triple_2}. However, we start observing that the resonance part becomes more relevant as the neutrino energies are larger. In Figs.~\ref{MicroBooNE_Triple_1} and~\ref{MicroBooNE_Triple_2}, the agreement with data is rather good although some overestimation can be observed at very forward angles. This can be due to the absence of some nuclear-medium effects in the SuSAv2-QE models related to binding energy effects at very low kinematics that can be corrected including additional corrections from RMF models at that particular kinematics. This is not the case in Fig.~\ref{MicroBooNE_Triple_3}, where a good agreement with the data is obtained at neutrino energies slightly higher than 1 GeV. Nevertheless, in Fig.~\ref{MicroBooNE_Triple_4} we tend to underestimate the experimental data, which could be ascribed to some lack of strength in our inelastic contributions. As shown in~\cite{PhysRevD.108.113008}, where we compared with the low-energy flux-averaged MINERvA data at similar kinematics (neutrino energy around 3.5 GeV) for a hydrocarbon target, we tend to underpredict the data possibly due to the lack of strength in the resonance channel and/or in other inelastic channels with respect to the results shown by other models used in generators. We should also note that in Fig.~\ref{MicroBooNE_Triple_4}, the cross section becomes negative for certain bins, which is an effect related to the application of the smearing matrix to both theoretical and experimental results, as detailed by MicroBooNE ~\cite{microboonecollaboration2024measurementthreedimensionalinclusivemuonneutrino}. In general, our global $\chi^{2}$ value (740.1) for the four sets of graphs and the individual values for each set are similar to the results given by other models~\cite{microboonecollaboration2024measurementthreedimensionalinclusivemuonneutrino} used in the experimental generators. 

\begin{figure*}[!htbp]
  \includegraphics[width=\textwidth]{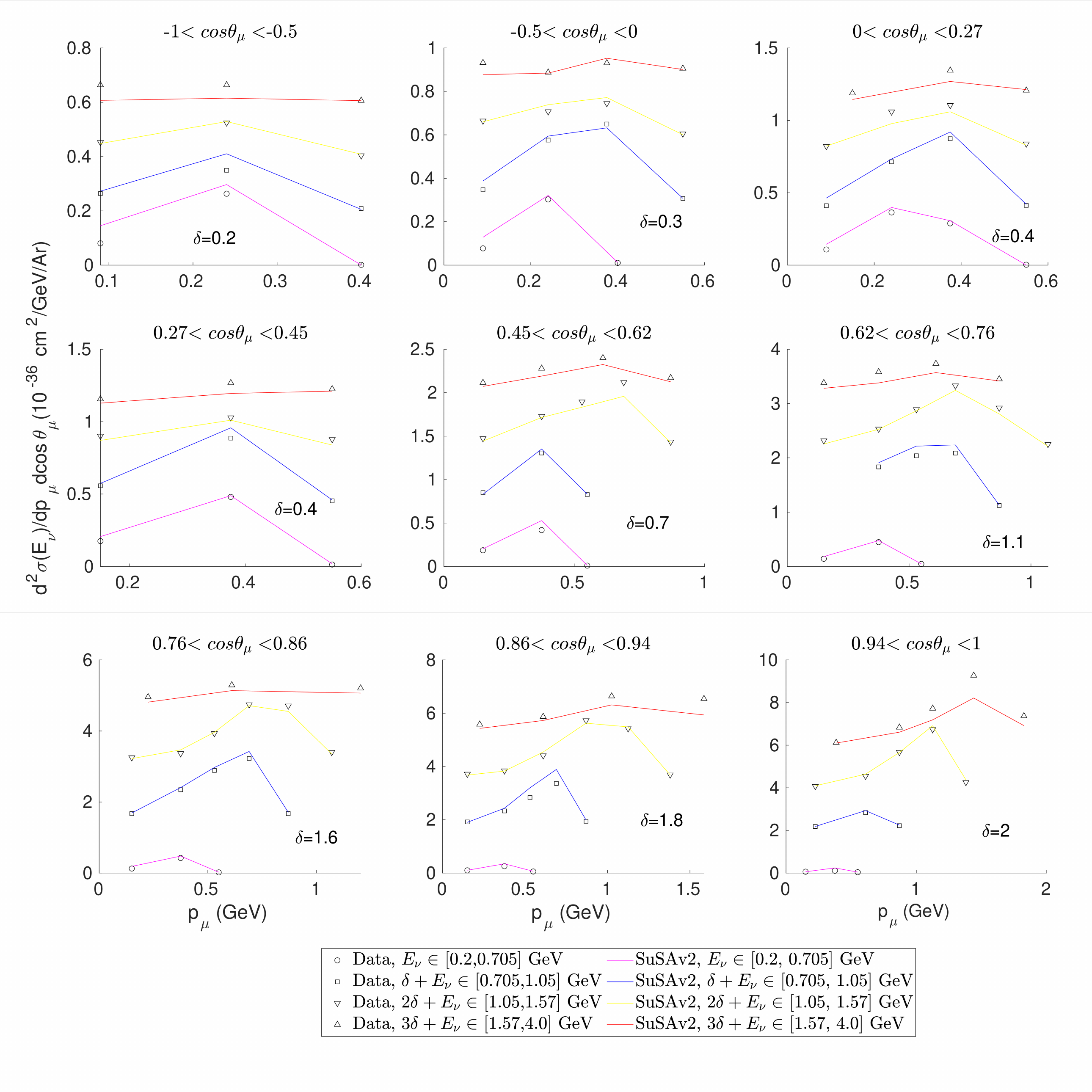}
    \caption{MicroBooNE CC inclusive flux-average differential cross section on $^{40}$Ar per nucleus. The data and the prediction  are shown
within each angular bin for each of the previous $E_{\nu}$ bins where, following the experimental paper, an offset has been introduced to avoid overlapping between the results and ease their visualization. The magnitude of the offset $\delta$, given in the same units of $10^{-36} cm^{2}/GeV/Ar$, is indicated at the bottom of each plot. Data from \cite{microboonecollaboration2024measurementthreedimensionalinclusivemuonneutrino}. \label{MicroBooNE_Triple_delta} }
\end{figure*}

In Fig.~\ref{MicroBooNE_Triple_costh}, we show the flux-averaged differential cross section in terms of the cosine of the muon scattering angle for each neutrino energy bin, which summarizes the information from Figs.~\ref{MicroBooNE_Triple_1}-\ref{MicroBooNE_Triple_4}. As expected, we get good agreement with data for the first three panels with some slight overestimation at lower neutrino energies, underpredicting on the contrary some data points on the last panel. This is consistent with the results shown in the previous plots.
In general, our global $\chi^{2}$ value (289.5) is larger than the one shown by other models~\cite{microboonecollaboration2024measurementthreedimensionalinclusivemuonneutrino} used in the experimental generators. In contrast, the individual values of $\chi^{2}$ for each panel, obtained using the covariance submatrixes, are rather similar to the other models in the generators, suggesting that the increase in the global value of $\chi^{2}$ comes from the correlation within the bins of the different panels.  For completeness, in Fig.~\ref{MicroBooNE_Triple_delta} we use the same format as shown in the original article~\cite{microboonecollaboration2024measurementthreedimensionalinclusivemuonneutrino} for Figs.~\ref{MicroBooNE_Triple_1}-\ref{MicroBooNE_Triple_4} which eases the comparison of our predictions with the NuWro results. 

\section{Conclusions}\label{Conclusion}

This paper presents a comparison of the SuSAv2 approach considering all reaction channels with recent NOvA and MicroBooNE CC inclusive neutrino cross section measurements. For the resonance regime, the Osaka DCC prescription for single-nucleon resonant structure functions has been applied to our framework. This model was already implemented in the SuSAv2 approach and validated in a previous work~\cite{PhysRevD.108.113008}. The other channels considered in the SuSAv2 framework were extensively analyzed and validated against electron and neutrino data in previous works~\cite{megias_inclusive_2016,megias_charged-current_2016,gonzalez-rosa_susav2_2022}.

In the case of the NOvA analysis, we have found an overall good agreement with the electron neutrino measurements, apart from some minor underestimation at very forward angles and large electron energies. On the contrary, the study of muon neutrino data have shown significant discrepancies with data mainly at forward angles, which can be related to our description of the inelastic channels. Note that for the muon neutrino case, the long tail of the flux allows us to explore larger kinematics than with the electron neutrino one, thus making the inelastic channels more relevant.  This conclusion was also drawn in a previous work~\cite{PhysRevD.108.113008} in which MINERvA data also were underestimated at similar kinematics. Note that NOvA tunes the magnitude and shape of the 2p2h contributions in their Monte Carlo generators to get closer to data, although this can worsen the agreement of the 2p2h models with electron scattering data and other neutrino measurements such as CC0$\pi$ T2K or MINERvA measurements. This seems to indicate that the source of these discrepancies could be connected with the inelastic channels.



Regarding our predictions at MicroBooNE kinematics, we tend to underestimate the total and single differential cross sections from~\cite{PhysRevLett.128.151801}, which contradicts previous analysis of CC-inclusive MicroBooNE and T2K data at similar kinematics~\cite{gonzalez-rosa_susav2_2022,PhysRevD.108.113008}. For MicroBooNE, where the mean neutrino energy is below 1 GeV, the dominant channel should be the QE one, as mostly observed. In that case, the possible lack of strength presented by our model in the inelastic channel is not expected to explain these discrepancies.  
In contrast, when analyzing the MicroBooNE double and single differential cross sections for different energy bins, published in \cite{microboonecollaboration2024measurementthreedimensionalinclusivemuonneutrino}, we observe that for the neutrino energy bins below 1.6 GeV the agreement with data is overall good, while some underestimation emerges at neutrino energies larger than 1.6 GeV, where the magnitude of the cross section is small in comparison with the lower $E_\nu$ region. Only for these particular bins of large neutrino energies, the discrepancies with data could be due to
limitations of our modeling in the description of the inelastic channels.




In general, we can conclude that the discrepancies with the NOvA data could be connected to our description of the inelastic channels at larger energies, but the MicroBooNE case is more difficult to explain, as we got a good agreement with previous MicroBooNE data and also T2K ones which explore similar kinematics. Further studies with the forthcoming MicroBooNE measurements will help to clarify this issue. Moreover, the exploration of new ingredients to be added to our inelastic models, such as other approaches for the inelastic structure functions, would be necessary to draw more definite conclusions about the source of these discrepancies. This could also be analyzed to address the discrepancies observed at very high energies for different experiments.


It is also important to note that with the ongoing implementation of the SuSAv2 and RMF models in GENIE and NEUT, which has been partially done for the QE and 2p2h contributions~\cite{PhysRevD.101.033003, PhysRevD.103.113003,CLAS:2021neh} and will be soon finished for the inelastic channels~\cite{Gardiner_inpreparation}, we will be able to combine and complement SuSAv2 predictions with other theoretical approaches and features present in the Monte Carlo generators, also leveraging their capabilities. These developments will allow us to improve consistency between the semi-inclusive and CC0$\pi$ predictions and the CC-inclusive ones provided by the SuSAv2 and RMF models.



\begin{acknowledgments}
This work was partially supported by the Spanish Ministerio de Ciencia, Innovación y Universidades and ERDF (European Regional Development Fund) under contracts PID2020-114687GB-100 and PID2023-146401NB-I00, by the Junta de Andalucia (grants No.~FQM160 and SOMM17/6105/UGR), by University of Tokyo ICRR’s Inter-University Research Program FY2023 (Ref.~2023i-J-001) $\&$ FY2024 (Ref.~2024i-J-001),  by the INFN under project Iniziativa Specifica NucSys and the University of Turin under Project BARM-RILO-23 (M.B.B.). J.G.R. was supported by a Contract PIF VI-PPITUS 2020 from the University of Seville (Plan Propio de Investigación y Transferencia). 
We also thank Steven Gardiner, Afroditi Papadopoulou, Benjamin Bogart, London Cooper-Troendle, Bannanje Nitish Nayak, Wenqiang Gu, Xin Qian  for fruitful discussions about statistical analysis related to the use of smearing matrices and covariance matrices for the comparison with MicroBooNE data.
\end{acknowledgments}

\bibliography{DCC.bib}

\end{document}